\documentclass[12pt]{article}
\usepackage[utf8]{inputenc}
\usepackage{amsmath,amsfonts,bbm}
\usepackage{setspace, comment}
\usepackage{pdfpages}
\usepackage{booktabs}
\usepackage{subcaption} % for subfigures
\usepackage[margin=1in]{geometry}
\usepackage{microtype}
\usepackage{graphicx}
\usepackage{natbib}

\makeatletter\let\c@figure\c@table\makeatother
\title{Dynamic Population Models with Temporal Preferential Sampling to Infer Phenology}
\author{Michael R. Schwob, Mevin B. Hooten, Travis McDevitt-Galles} % comment to remove personal information
\date{}

\begin{document}

\maketitle

\section*{Abstract}
To study population dynamics, ecologists and wildlife biologists use relative abundance data, which are often subject to temporal preferential sampling. Temporal preferential sampling occurs when sampling effort varies across time. To account for preferential sampling, we specify a Bayesian hierarchical abundance model that considers the dependence between observation times and the ecological process of interest. The proposed model improves abundance estimates during periods of infrequent observation and accounts for temporal preferential sampling in discrete time. Additionally, our model facilitates posterior inference for population growth rates and mechanistic phenometrics. We apply our model to analyze both simulated data and mosquito count data collected by the National Ecological Observatory Network. In the second case study, we characterize the population growth rate and abundance of several mosquito species in the \textit{Aedes} genus.

\vspace{.2cm}

\noindent \textbf{Key words:} abundance, ignorability, mechanistic modeling, phenometrics, time series

\section{Introduction}

The timing of ecological events is a fundamental question in the field of phenology \citep{forrest2010toward}. Conventional approaches to studying phenology involve summary statistics, or phenometrics, that capture the timing of cyclical biological events \citep{demidova2021synthesis}. Phenometrics for population dynamics may include the first or last date of positive population growth \citep{anderson2013modeling,bewick2016resource}. By contrast, ecologists commonly use hierarchical temporal models to learn about how and why population size changes over time \citep{potts2006comparing}. We seek to make both phenological and ecological inference on population dynamics using modern hierarchical Bayesian models that allow for embedded mechanistic and observation processes. We specify the model in such a way to extract relevant phenometrics and associated uncertainties as derived quantities based on our defined mechanistic process (i.e., the latent population growth rate).

Ecologists working in temperate environments tend to design field studies that align with the growing season \citep{pau2011predicting}. The resulting data sets are collected using preferential sampling designs, which may induce nonignorable missingness \citep{zachmann2022bayesian}. In the presence of temporal preferential sampling, inference during periods of infrequent observation will be biased \citep{monteiro2019modelling}. We adapt preferential sampling approaches for the discrete temporal domain to improve posterior inference on the abundance process during periods of infrequent observation.

Remedies for spatial preferential sampling have been more popular than those for temporal preferential sampling in ecology \citep{watson2021perceptron}. However, phenological studies may be subject to temporal preferential sampling, including frequent and short gaps \citep{zhang2009sensitivity}, fewer and longer gaps \citep{rolevcek2007sampling}, and asynchronous and inconsistent gaps in data collection \citep{cole2012mind}. Records of population dynamics have infrequent and long gaps in observation because the data are collected seasonally. However, the existing remedies to temporal preferential sampling focus solely on frequent and short gaps in observations \citep{karcher2016quantifying,monteiro2019modelling,monteiro2020modelling}. Furthermore, existing temporal preferential sampling remedies are only applicable in the continuous-time domain, whereas most abundance studies collect counts for discrete-time population dynamics. In studies that use hidden Markov models, temporal preferential sampling is referred to as state-dependent missingness. Several recent algorithms accommodate state-dependent missingness in observations of homogeneous Markov chains \citep{speekenbrink2021ignorable,hoskovec2022infinite}; however, the existence of seasonal variation in population dynamics precludes the use of these algorithms. Therefore, existing solutions to temporal preferential sampling and state-dependent missingness are not applicable to phenological or abundance data sets. We account for discrete-time temporal preferential sampling in the form of a probit regression that is suitable for many phenological and abundance data sets that exhibit seasonal variation.

We propose a hierarchical Bayesian abundance model in section 2 that characterizes population dynamics using relative abundance data from irregular surveys. We specify the process model to characterize the population growth rate and infer phenometrics, and we modify the data model to account for temporal preferential sampling. In section 3, we demonstrate the proposed model along with its non-preferential variant on simulated data and a mosquito case study with data collected by the National Ecological Observatory Network (NEON). The data collection associated with the NEON study involved temporal preferential sampling. We seek to determine if posterior inference resulting from a preferential and a non-preferential model differs under the NEON sampling protocol. We also show that the bias varies when interpolating population dynamics during infrequently observed periods. In section 4, we conclude with a discussion of mechanistic modeling and temporal preferential sampling for phenology and population dynamics.

\section{Methods}

We let $T$ denote the number of days throughout the study period and $J$ denote the number of species in the study. We also let $\mathbf{y}(t)\equiv(y_1(t),...,y_J(t))'$ and $\boldsymbol{\lambda}(t)\equiv(\lambda_1(t),...,\lambda_J(t))'$ denote the observed and true abundance of all species on day $t\in\{1,...,T\}$ and aim to recover the latent abundance process $\boldsymbol{\Lambda}\equiv(\boldsymbol{\lambda}(1),...,\boldsymbol{\lambda}(T))$ using temporal state-space modeling. The specification of the observation distribution $[\mathbf{y}(t)|\boldsymbol{\lambda}(t)]$ and the process distribution $[\boldsymbol{\lambda}(t)|\boldsymbol{\lambda}(t-1)]$ are fundamental to obtaining interpretable inference in temporal state-space models. We denote the data model as $\mathbf{y}(t)=\mathcal{H}(\boldsymbol{\lambda}(t);\boldsymbol{\Theta}_d;\boldsymbol{\epsilon}(t))$, where $\mathcal{H}$ is an observation function that maps the latent process $\boldsymbol{\lambda}(t)$ to the data with noise process $\boldsymbol{\epsilon}(t)$. Similarly, we can write the latent process evolution model as $\boldsymbol{\lambda}(t)=\mathcal{M}(\boldsymbol{\lambda}(t-1);\boldsymbol{\Theta}_e;\boldsymbol{\eta}(t))$, where $\mathcal{M}$ is the evolution operator and $\boldsymbol{\eta}(t)$ is a noise process \citep{wikle2010general}. In general, the noise processes may be Gaussian or non-Gaussian. We assume a first-order Markov process for the evolution model, which is common in ecological studies of population dynamics \citep{usher1979markovian}. 

\subsection{The Process Model}

We model abundance dynamically and focus on specifications involving mechanistic population models. A common mechanistic approach to modeling population dynamics is to specify the latent process evolution as 
\begin{equation}\label{eq:ross}
    \mathcal{M}(\boldsymbol{\lambda}(t-1);\boldsymbol{\Theta}_e;\boldsymbol{\eta}(t)) = \exp\left(\mathbf{A}\log\boldsymbol{\lambda}(t-1) + \mathbf{B}\mathbf{x}(t) + \boldsymbol{\eta}(t)\right),
\end{equation}
where $\boldsymbol{\Theta}_e\equiv\{\mathbf{A},\mathbf{B}\}$, $\boldsymbol{\eta}(t)\sim\text{N}(\boldsymbol{0},\sigma^2\mathbf{I})$, and $\mathbf{x}(t)\equiv(x_1(t),...,x_p(t))'$ are species-independent covariates of interest at time $t$ \citep{ross2015combined}. In the multivariate setting, we let $\exp(\cdot)$ and $\log(\cdot)$ be element-wise operations over vectors. We let $\mathbf{A}\equiv\text{diag}(\alpha_1,...,\alpha_J)$ and $\mathbf{I}$ be the $J\times J$ identity matrix. Species $j$ experiences density dependence when $\alpha_j<1$, positive association with population density when $\alpha_j>1$, and density independence when $\alpha_j=1$ \citep{dennis2006estimating}. We let $\mathbf{B}\equiv(\boldsymbol{\beta}_1,...,\boldsymbol{\beta}_J)'$, where $\boldsymbol{\beta}_j\equiv(\beta_{j,1},...,\beta_{j,p})$ are the time-invariant coefficients of covariates $\mathbf{x}$ for species $j$. The process model in (\ref{eq:ross}) can be written as
\begin{equation}\label{eq:init}
    \log\boldsymbol{\lambda}(t) \sim \text{N}\left(\mathbf{A}\log\boldsymbol{\lambda}(t-1) + \mathbf{B}\mathbf{x}(t),\sigma^2\mathbf{I}\right),
\end{equation}
where the species-level parameters $\boldsymbol{\beta}_j$ convey how the abundance of species $j$ responds to a change in environment and resources. 

Most phenology studies focus on species with population dynamics that experience a high degree of seasonality. We modify the process model in (\ref{eq:init}) for species with population dynamics that are intrinsically tied to the environment and, therefore, seasonal. The more seasonality influences population dynamics, the more the dynamic is centered on environmental trends. We account for the seasonality of population dynamics by anchoring the log-abundance using the environmental trend $\mathbf{B}\mathbf{x}(t)$ such that
\begin{equation}
    \log\boldsymbol{\lambda}(t) - \mathbf{B}\mathbf{x}(t) \sim \text{N}\left(\mathbf{A}(\log\boldsymbol{\lambda}(t-1) - \mathbf{B}\mathbf{x}(t-1)),\sigma^2\mathbf{I}\right),
\end{equation}
where $|\alpha_j|<1$ because the population is assumed to be density dependent \citep{royama2012analytical}; this provides a trend-stationary and autoregressive stochastic model for population dynamics that are heavily influenced by the environment \citep{shumway2000time}. Thus, our process model for log-abundance can be expressed as
\begin{equation}
    \log\boldsymbol{\lambda}(t) \sim \text{N}\left(\mathbf{B}\mathbf{x}(t) - \mathbf{A}\mathbf{B}\mathbf{x}(t-1) + \mathbf{A}\log\boldsymbol{\lambda}(t-1),\sigma^2\mathbf{I}\right),\label{eq:logabundance}
\end{equation}
for $t=2,...,T$. When $t=1$, we assume $\log\boldsymbol{\lambda}(1)\sim \text{N}(\boldsymbol{\mu}_1,\sigma^2_1\mathbf{I})$. Although trend-stationary stochastic processes are commonly found in econometric studies, they are underutilized in population dynamic studies for species with seasonal abundance processes. When applicable, accounting for trend stationarity improves the stability and performance of MCMC algorithms \citep{mcculloch1994bayesian}.

The Gompertz form of density dependence has performed well in various population dynamic studies \citep{dennis2006estimating,knape2012patterns}. Our process model implies stochastic Gompertz population growth, which can be written as the heterogeneous Malthusian growth function $\boldsymbol{\lambda}(t)=(\mathbf{I}+\text{diag}(\mathbf{g}(t)))\boldsymbol{\lambda}(t-1),$ where $\mathbf{g}(t)\equiv(g_1(t),...,g_J(t))'$. The species-level per capita growth rate on day $t$ can be written as
\begin{equation}\label{eq:growth}
    g_j(t)=\exp\left(\boldsymbol{\beta}_j(\mathbf{x}(t)-\alpha_j\mathbf{x}(t-1))\right)\lambda_j(t-1)^{\alpha_j-1}e^{\epsilon_{j,t}}-1,
\end{equation}
where $\epsilon_{j,t}\sim \text{N}(0,\sigma^2)$. We consider the posterior median of the log-normal distributed random variable $e^{\epsilon_{j,t}}$, which assumes an absolute loss function; we assume this loss function because it is less sensitive to skewness resulting from exponentiating the log-abundance process. Additionally, the posterior median is more efficient and robust than the posterior mean for log-normally distributed random variables with $\sigma>0.546$, which is true in our case studies \citep{zellner1971bayesian,rao2016bayesian}. Thus, our derived phenometric is 
\begin{equation}\label{eq:pm_growth}
    \psi \equiv \min\bigg\{ t \in\{1,...,T\} : \exp\left(\boldsymbol{\beta}_j(\mathbf{x}(t)-\alpha_j\mathbf{x}(t-1))\right)\lambda_j(t-1)^{\alpha_j-1}-1 > 0\bigg\},
\end{equation}
which is the first day that the posterior median of $g_j(t)$ is positive. We use posterior densities to quantify uncertainty for the derived phenometric $\psi$, which is not possible using conventional phenological approaches. We present the derivation of $g_j(t)$ and its median in Appendix A.

\subsection{The Data Model}

For the observation function $\mathcal{H}$, we specify
\begin{equation}
    y_j(t)\sim\text{Pois}(\lambda_j(t)\cdot\omega(t)),
\end{equation}
where $\omega(t)$ captures heterogeneity in sampling effort. The conditional Poisson data model is a common choice in temporal abundance modeling \citep{hooten2019bringing}. For invertebrate trapping studies, we specify $\omega(t)=w(t)\cdot h(t)$, where $w(t)$ is the proportion of the trap sampled and $h(t)$ is the sampling duration of the trap standardized to a single day length in hours. If sampling effort was constant across traps, we set $\omega(t)=1, \; \forall t$.

To account for temporal preferential sampling, we introduce new notation and indices for time. We let $N$ denote the number of days within the study period and $n$ denote the number of observed days. Further, we let $t_i$ denote the $i$th day of the study and $\tilde{t}_k$ denote the $k$th observed day of the study. We define $\mathcal{T}$ as the set of observed days, such that
\begin{align}
    \mathcal{T} \equiv \{\tilde{t}_1,\tilde{t}_2,...,\tilde{t}_n\} \subseteq \{t_1,t_2,...,t_N\}.
\end{align}
In the event that every day is observed, temporal preferential sampling is not present and $\mathcal{T}=\{t_1,t_2,...,t_N\}$. We let $\boldsymbol{\tau}\equiv\left(\tau_1,\tau_2,...,\tau_N\right)'$ be a binary vector of length $N$ such that
\begin{align}
    \tau_i = \begin{cases} 1, \quad t_i \in \mathcal{T}\\
    0, \quad t_i \not\in \mathcal{T}
    \end{cases}.
\end{align}
Finally, we let $\boldsymbol{\lambda}(t_i)$ denote the abundance processes for all species at time $t_i$. The abundance processes exist throughout the entire study regardless of which days were observed.

We express the joint distribution of the underlying process $\boldsymbol{\Lambda}$, observation indicators $\boldsymbol{\tau}$, and the observations of the underlying process $\mathbf{Y}\equiv(\mathbf{y}(1),...,\mathbf{y}(n))$ as
\begin{equation}
    [\boldsymbol{\Lambda},\boldsymbol{\tau},\mathbf{Y}]=[\mathbf{Y}|\boldsymbol{\tau},\boldsymbol{\Lambda}][\boldsymbol{\tau}|\boldsymbol{\Lambda}][\boldsymbol{\Lambda}],
\end{equation}
where ``$[\cdot|\cdot]$" represents the conditional probability density or distribution function \citep{gelfand1990sampling}. We condition $\boldsymbol{\tau}$ on $\boldsymbol{\Lambda}$ because the observation of days depends on the process of interest. The observation time model is usually specified as a discretized continuous-time inhomogeneous Poisson process, which is ideal for continuous time series with many short gaps of missing data \citep{karcher2016quantifying,monteiro2019modelling}. However, it is more common that there are fewer, longer gaps of missing data in discrete-time population dynamic studies. Therefore, we specify $[\boldsymbol{\tau}|\boldsymbol{\Lambda}]$ for discrete-time abundance counts that experience long, seasonal gaps in data collection. We consider the following distribution of observed days conditioned on the process of interest:
\begin{equation}\label{eq:taueq}
  [\tau_i|\boldsymbol{\theta},\boldsymbol{\lambda}(t_i),\tilde{\lambda}]\equiv\text{Bern}(p(t_i))  
\end{equation}
with the probit link function
\begin{align}\label{eq:probit}
    p(t_i) &= \Phi\left(\theta_0 + \theta_1 \mathbbm{1}_{\left\{\boldsymbol{\lambda}(t_i)'\mathbf{1} \ge \tilde{\lambda}\right\}}\right),
\end{align}
where $\Phi(\mu)$ is the standard normal cumulative distribution function evaluated at $\mu$ and  $\boldsymbol{\theta}\equiv(\theta_0, \theta_1)'$ are probit regression coefficients. The indicator function $\mathbbm{1}_{\{\cdot\}}$ evaluates to 1 when the number of observed mosquitoes exceeds the threshold $\tilde{\lambda}$, and 0 otherwise. We specify a prior for the threshold as $\tilde{\lambda}\sim\text{Ga}(\alpha_\lambda,\beta_\lambda)$.

Probit regression is a common regression technique for analyzing binary ecological data \citep{trexler1993nontraditional,hooten2019bringing}. We consider probit regression because we can induce dependence of the binary vector $\boldsymbol{\tau}$ on the abundance process using data augmentation \citep{albert1993bayesian}. Thus, we introduce the latent variables $z(t_i)$ and model $[\tau_i|\boldsymbol{\theta},\boldsymbol{\lambda}(t_i),\tilde{\lambda}]$ using
\begin{align}
    \tau_i &= \begin{cases}1, & z(t_i) > 0\\ 0, & z(t_i) \le 0\end{cases},\label{eq:tps1}\\
    z(t_i) &\sim \text{N}\left(\theta_0 + \theta_1 \mathbbm{1}_{\left\{\boldsymbol{\lambda}(t_i)'\mathbf{1} \ge \tilde{\lambda}\right\}},1\right),\label{eq:tps2}
\end{align}
which implies the same model as (\ref{eq:taueq})-(\ref{eq:probit}); however, the specification in (\ref{eq:tps1})-(\ref{eq:tps2}) results in conjugate updates for both the latent variables $z(t_i)$ and $\boldsymbol{\theta}$ assuming a multivariate normal prior on $\boldsymbol{\theta}$. The observations $\mathbf{Y}$ are no longer assumed to be conditionally independent because we explicitly define the data model to account for the dependence of $\boldsymbol{\tau}$ on $\boldsymbol{\Lambda}$. To complete the data model, we specify $y_j(\tilde{t}_k)\sim\text{Pois}(\lambda_j(\tilde{t}_k)\cdot\omega(\tilde{t}_k))$ based on the observations in set $\mathcal{T}$.

Tests for temporal preferential sampling are underdeveloped \citep{watson2021perceptron}. However, $\theta_1$ in (\ref{eq:probit}) provides a convenient model-based quantity that can be used to assess discrete-time temporal preferential sampling. When sampling mechanisms favor periods of higher abundance, we expect $\theta_1>0$. Thus, we may test for the presence of temporal preferential sampling using posterior inference on $\theta_1$.

The full Bayesian hierarchical model is provided in Appendix B. We implemented the model in Julia to reduce computation time \citep{Julia-2017}; alternatively, it could be fit using standard MCMC software. We used conventional, conjugate priors for $\boldsymbol{\theta}$, $\boldsymbol{\mu}_\beta$, $\boldsymbol{\Sigma}_\beta$, $\boldsymbol{\alpha}$, $\sigma^2$, and $\tilde{\lambda}$ to facilitate computation.

\section{Application}

We analyzed two abundance data sets subject to temporal preferential sampling: A simulated data set and a terrestrial site observed by the National Ecological Observatory Network. To understand the effect of temporal preferential sampling on posterior inference, we compared the proposed preferential model to its non-preferential variant, which does not contain the preferential component in (\ref{eq:tps1})-(\ref{eq:tps2}). The discrepancy between the resulting inference indicates the effects of neglecting the presence of temporal preferential sampling. We implemented both case studies  using MCMC algorithms and 500,000 iterations with a burn-in of 200,000 iterations.

\subsection{Simulated Abundance}

We simulated a log-abundance process for a single species using the process model in (\ref{eq:logabundance}). With $J=1$, we let $\mathbf{A}=0.98$ be a scalar, $\mathbf{B}=(\beta_1,\beta_2)=(0.1, 0.3)$ be a row vector, and $\sigma^2=0.03$. For the environmental covariates $\mathbf{x}$, we used an intercept and growing degree day (GDD) estimates at Harvard Forest in Massachusetts, USA, from 2014-2016. We quantified GDD using estimated temperature from the Oregon State PRISM Climate data with a baseline temperature of $10^{\circ}$C and a cutoff temperature of $30^{\circ}$C \citep{neteler2011terra,PRISM}. We simulated the sequence of fully observed counts $\mathbf{y}^*\equiv(y(t_1),...,y(t_N))'$ using a Poisson distribution with abundance as the intensity; we let $\omega(\tilde{t}_k)=1$ because we assume homogeneous sampling effort across traps.

\begin{figure}[!h]
    \centering
    \caption{Abundance estimates under three sampling mechanisms. In (a)-(c), the true abundance and the observed abundance are depicted as black lines and dots, respectively. The red lines and polygons depict the posterior means and 95\% credible intervals under the non-preferential model. The blue lines and polygons depict the posterior means and 95\% credible intervals under the preferential model.}
    \includegraphics[width=\textwidth]{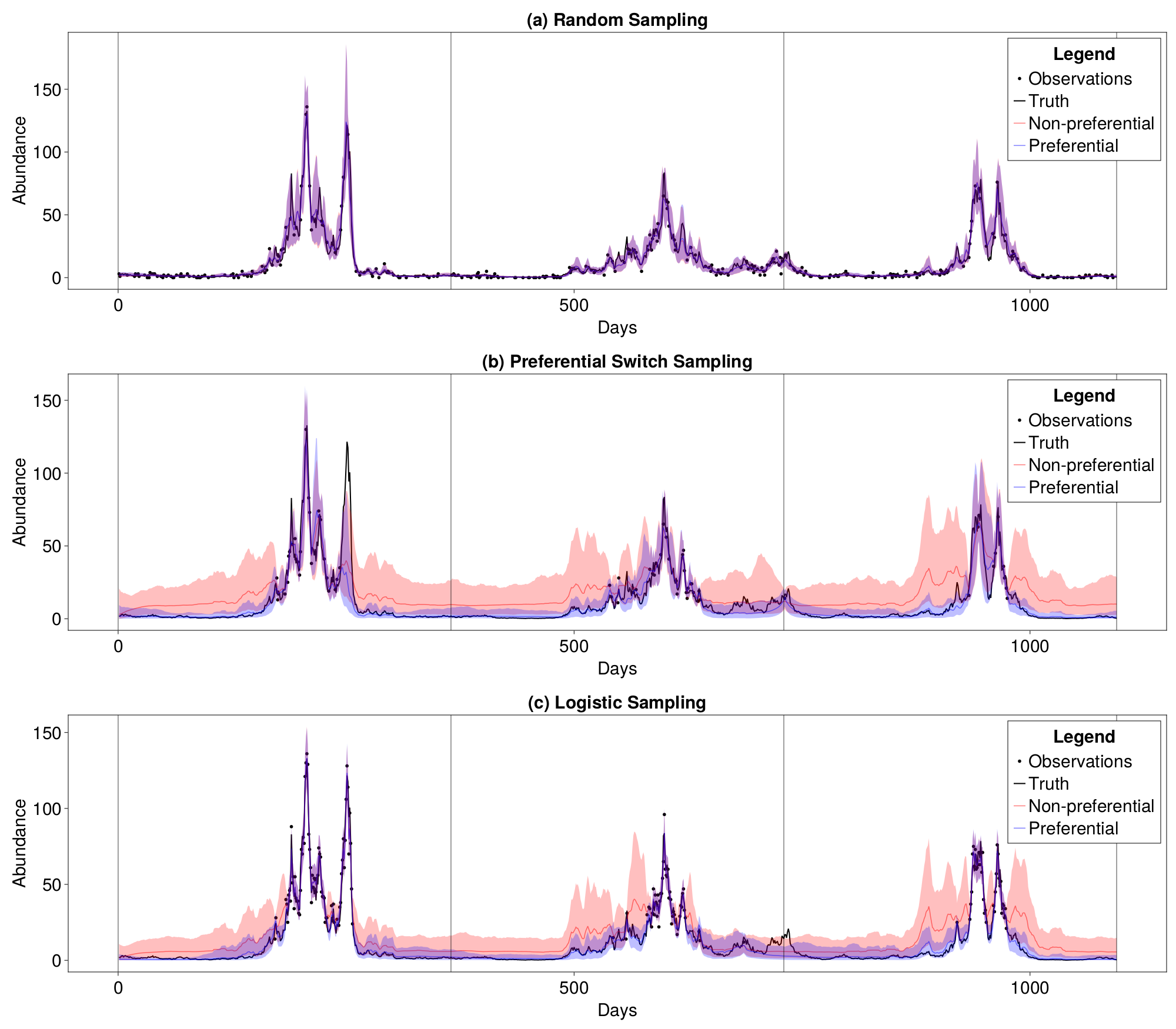}
    \label{fig:simEstimates}
\end{figure}

We considered three common sampling mechanisms found in abundance studies to determine which counts in $\mathbf{y}^*$ were retained for analysis: Random sampling, where each day in the study was observed with equiprobability (0.3); Preferential switch sampling, where each day that abundance exceeded a threshold (15) was observed with equiprobability (0.3) and each day that abundance did not exceed that threshold was unobserved; and logistic sampling, where the probability of observing day $t_i$ was $\text{logit}^{-1}(-10 + 0.4\cdot\lambda(t_i))$. We fit the preferential and non-preferential models to the retained observed counts to determine the effect of ignoring temporal preferential sampling on posterior inference. Observations and posterior estimates for abundance under the three scenarios are shown in Figure \ref{fig:simEstimates}. 

To quantify the error for each model in each scenario, we considered root-mean-squared error (RMSE) for the abundance process $\{\lambda(t_i)\}$, which is conventional in ecological model comparisons for species richness \citep{mouillot1999comparison}. Table \ref{tab:RMSE} contains the RMSE for the preferential and non-preferential models for the three sampling scenarios. 

\begin{figure}
    \centering
    \caption{Posterior estimates under the preferential and non-preferential model for random, preferential switch, and logistic sampling. True values are indicated by horizontal black lines.}
    \begin{subfigure}{0.45\textwidth}
        \includegraphics[width=\textwidth]{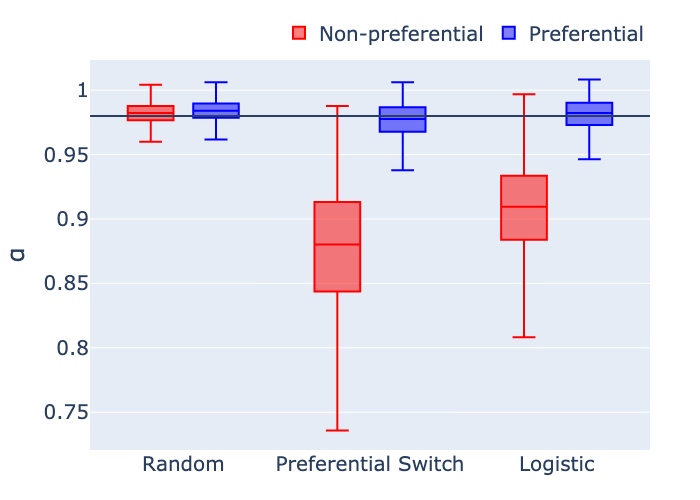}
        \caption{$\alpha$}
    \end{subfigure}
    \begin{subfigure}{0.45\textwidth}
        \includegraphics[width=\textwidth]{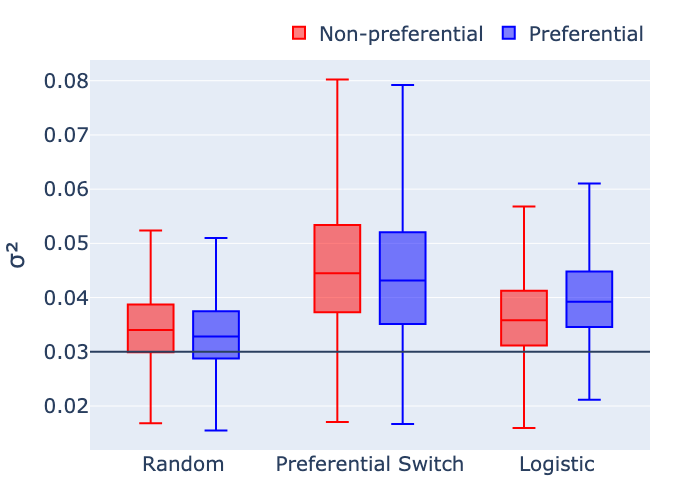}
        \caption{$\sigma^2$}
    \end{subfigure}
    \begin{subfigure}{0.45\textwidth}
        \includegraphics[width=\textwidth]{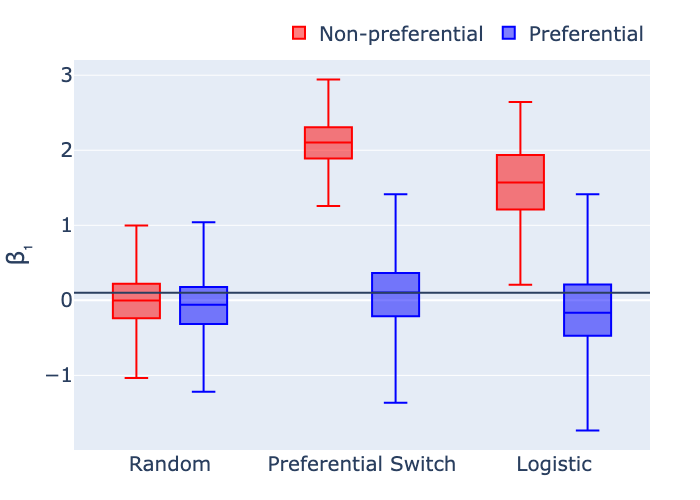}
        \caption{$\beta_1$}
    \end{subfigure}
    \begin{subfigure}{0.45\textwidth}
        \includegraphics[width=\textwidth]{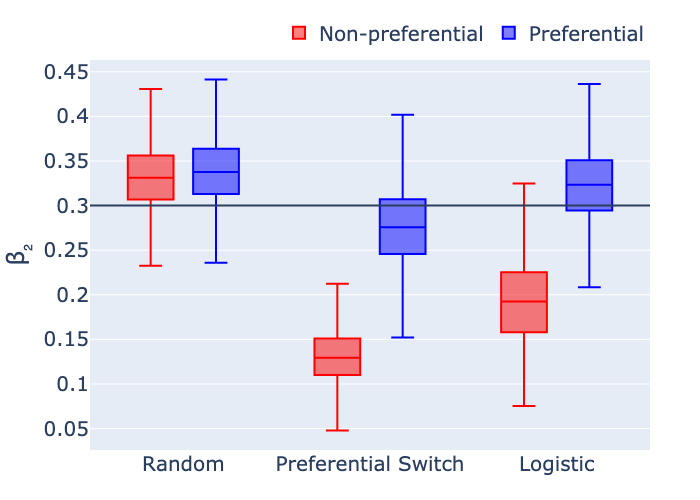}
        \caption{$\beta_2$}
    \end{subfigure}
    \label{fig:subFig}
\end{figure}

With random sampling, we did not expect the competing models to provide significantly different results because the sampling mechanism did not engage in temporal preferential sampling. Empirically, the addition of the preferential data model in (\ref{eq:tps1})-(\ref{eq:tps2}) did not improve posterior inference because the missingness in the data was ignorable. The posterior estimates for the parameters governing the log-abundance process were recovered by both models, as depicted in Figure \ref{fig:subFig}. The 95\% posterior credible interval for $\theta_1$ was (-0.53,0.46), which implies that temporal preferential sampling is not detected under the preferential model.

Under preferential switch sampling, abundance was overestimated using the non-preferential model during periods of infrequent observation, whereas the preferential model resulted in more accurate estimates during such periods. Additionally, the RMSE for the preferential model was lower than that for the non-preferential model. Both models adequately recovered $\sigma^2$. The posterior median for $\alpha$ under the preferential model was closer to the true value than under the non-preferential model. Finally, the 95\% posterior credible intervals for $\beta_1$ and $\beta_2$ contained the true value under only the preferential model.

For the logistic sampling scenario, the preferential model outperformed the non-preferential model because the observed days were highly concentrated during summers when abundance was notably high. During periods of infrequent observation, the estimated abundance from the preferential model was much closer to the truth than the estimates from the non-preferential model. Both models accurately recovered $\sigma^2$. Additionally, the posterior medians for the remaining parameters were close to the true values under the preferential model. The true values for $\alpha$ and $\beta_2$ were recovered under the non-preferential model; however, the 95\% credible interval for $\beta_1$ did not contain the true value.

In the presence of temporal preferential sampling, the preferential model resulted in more precise estimates for the abundance process throughout the entire study period, as evidenced by the narrower 95\% credible intervals in Figure \ref{fig:simEstimates}. The abundance process was recovered accurately and precisely under the preferential model during periods of infrequent observation, whereas the non-preferential model tended to overestimate abundance during these periods. Additionally, the true value for each parameter in the log-abundance process was captured under the preferential model, whereas $\beta_1$ and $\beta_2$ were poorly estimated under the non-preferential model for both preferential sampling scenarios. Thus, we obtained inadequate posterior inference via the non-preferential model in the presence of temporal preferential sampling.

\subsection{Mosquito Abundance}

We analyzed mosquito count data collected by NEON. The NEON mosquito monitoring program followed a standardized sampling protocol conducted across a broad geographical range and aimed to collect adult mosquito activity data for phenological modeling \citep{hoekman2016design,NEON}. NEON collected data throughout each year with two distinct sampling approaches based on levels of mosquito activity: An ``off-season” with infrequent sampling during periods of low abundance, and a ``field season," where sampling intensity increased concurrently with abundance. Because the decision to start and stop consistent data collection was motivated by the abundance process, NEON engaged in temporal preferential sampling. We sought to determine if posterior inference resulting from the preferential and non-preferential model differed under the NEON sampling protocol.

We expressed the environmental covariates as a convolution to account for latency in the reproductive and larval stages of the mosquito life cycle: 
\begin{equation}\label{eq:kernel}
    \mathbf{x}(t) = \int \mathbf{G}(\boldsymbol{\phi},t,\tilde{\tau})\tilde{\mathbf{x}}(\tilde{\tau})d\tilde{\tau},
\end{equation}
where $\tilde{\mathbf{x}}(\tilde{\tau})$ represents the original continuous-time environmental variables, $\mathbf{x}(t)$ is a smoothed version of $\tilde{\mathbf{x}}(\tilde{\tau})$ resulting from the convolution, and $\mathbf{G}$ is a diagonal matrix of basis function $g_l(\boldsymbol{\phi},t,\tilde{\tau})$ for $l=1,...,p$ associated with each covariate. The convolution in (\ref{eq:kernel}) is generalized to account for temporal lag in population dynamics \citep{aukema2008movement,chi2008spatial}. We constructed the convolution in the form of a backward moving average to account for the lagged effect of environmental conditions on the mosquito life cycle. If the basis functions are specified as 
\begin{equation}
    g_l(\phi,t,\tilde{\tau})=\begin{cases} \frac{1}{\phi}, & t-\phi < \tilde{\tau} < t\\
0, & \text{otherwise}\end{cases},
\end{equation}
then the population growth rate is a function of the average environmental conditions over the past $\phi$ days. The parameter $\phi$ can be specified using expert knowledge of mosquito biology or estimated in the model when enough data exist to identify it in addition to other model parameters. We used a 14-day backward moving average of GDD at each site throughout the study period because GDD has been reported to influence mosquito dynamics \citep{field2019satellite}; the 14-day time frame was chosen to account for the stages in the mosquito life cycle preceding sexual maturity \citep{crans2004classification}.

We restricted analysis to the \textit{Aedes} genus, in which several species are vectors for dengue fever, yellow fever, and the Zika virus \citep{suwanmanee2017dengue}. Of the NEON terrestrial sites where the \textit{Aedes} genus was present, we analyzed counts at the University of Notre Dame Environmental Research Center (UNDE) because the counts were obtained daily during the field season. UNDE observed counts for three species in the \textit{Aedes} genus: \textit{Aedes canadensis}, \textit{Aedes excrucians}, and \textit{Aedes punctor}. We analyzed counts from 2016 to 2019 in this study.

\begin{figure}
    \centering
    \caption{\textit{Top:} Posterior IQR and median for the abundance process of \textit{Aedes punctor} at the UNDE site from 2016-2019. Black dots denote observed abundances, lines denote the posterior medians, and polygons denote the posterior IQRs. \textit{Bottom:} Derived posterior means for the growth rate of \textit{Aedes punctor} at the UNDE site from 2016-2019. The blue lines and polygons correspond to inference provided by the preferential model, whereas the red corresponds to the non-preferential model.}
    \includegraphics[width=\textwidth]{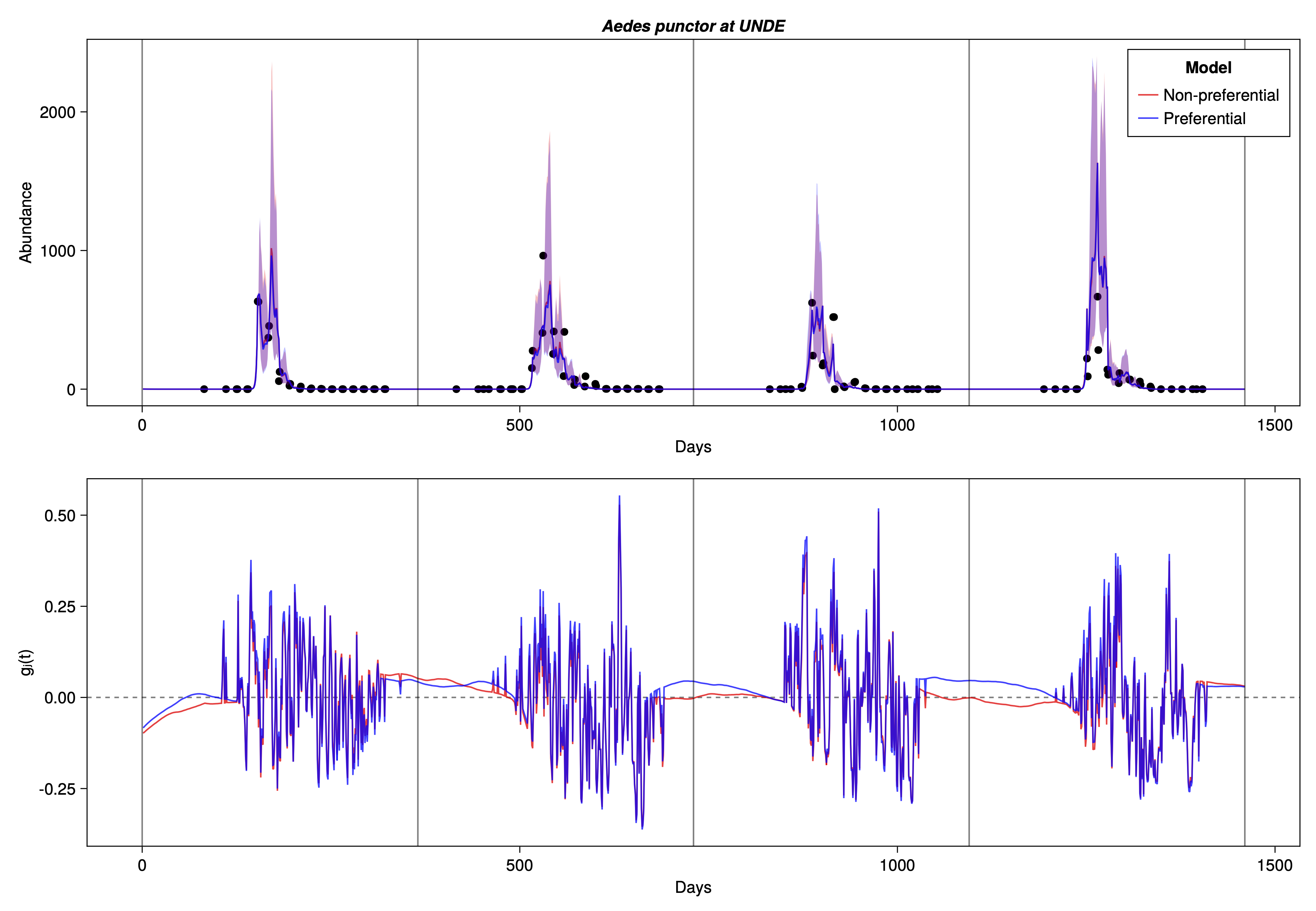}
    \label{fig:estimates_3}
\end{figure}

We considered an absolute loss function when providing inference because it is robust to skewness that may result when exponentiating the log-abundance process. Therefore, we reported posterior medians for the abundance process of \textit{Aedes punctor} in Figure \ref{fig:estimates_3}; the analogous plots for \textit{Aedes canadensis} and \textit{Aedes excrucians} can be found in Supplemental Figures 1 and 2. We also reported posterior interquartile ranges (IQR) to accompany the posterior point estimates. As expected, some observed counts exceed the posterior IQR due to the inflated counts resulting from extended trapping duration. Similarly, we computed population growth rates by considering the posterior median of $\mathbf{g}(t)$ in (\ref{eq:growth}), which depends on $\boldsymbol{\alpha}$ and $\mathbf{B}$. The estimated growth rates are also provided with each species' abundance process.

We inferred the first day in which a species has positive population growth by computing the derived phenometric $\psi$. The posterior estimates of $\psi$ are depicted in Supplemental Figure 3 for 2017-2019 alongside the same phenometric derived using conventional summary statistics \citep{jones2018herbarium}. Many phenological approaches cannot provide uncertainty quantification, whereas our mechanistic hierarchical model formally allowed us to quantify uncertainty with $\psi$ via posterior densities. 

The posterior inference provided by the preferential and non-preferential models matched throughout the entire study period for each species. Neither model overestimated the relatively low mosquito abundance during the winter, a behavior evident in the non-preferential model in the simulated case study. The resulting inference from both models was similar because NEON consistently obtained enough zero counts preceding the annual rise and following the annual decline in mosquito populations; fitting these zero-counts to the non-preferential model resulted in adequate estimates for $\boldsymbol{\alpha}$ and $\mathbf{B}$. Although NEON explicitly engaged in temporal preferential sampling, their sampling mechanism mitigated the effect that temporal preferential sampling had on posterior inference by sampling infrequently during the off-season until a mosquito is detected. 

% In the preferential model, the posterior mean for $\tilde{\lambda}$ was 1.24. Thus, the NEON off-season was estimated to occur prior to the detection of mosquitoes, which aligns with the true NEON sampling protocol.

\begin{figure}
    \centering
    \caption{\textit{Top:} Posterior IQR and median for the abundance process of \textit{Aedes punctor} at the UNDE site when removing zero-count observations. Full black dots denote observed abundances, black circles denote removed zero-count observations, lines denote the posterior medians, and polygons denote the posterior IQRs. \textit{Bottom:} Derived posterior means for the growth rate of \textit{Aedes punctor} at the UNDE site when removing zero-count observations.}
    \includegraphics[width=\textwidth]{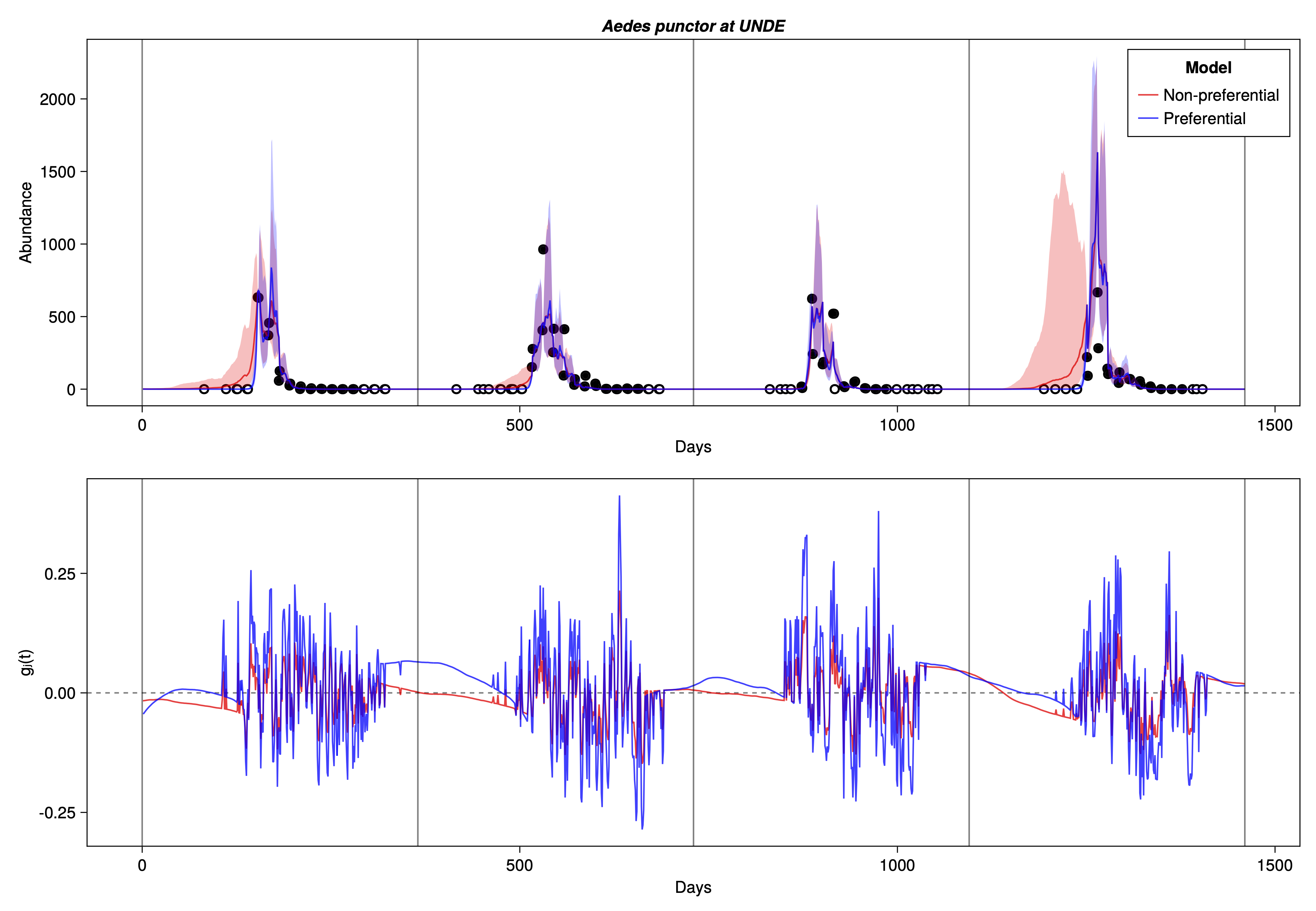}
    \label{fig:wo0}
\end{figure}

We also analyzed a subset of the UNDE data, where observations were removed for days in which zero mosquitoes were detected; we refer to this subset as the second scenario, which resembles many abundance studies that only experience positive counts \citep{turchin2013complex}. The posterior inference under the competing models significantly differed under the second scenario, as seen in Figure \ref{fig:wo0}. The first observed counts of \textit{Aedes punctor} were positive in 2016, 2017, and 2019. Consequently, abundance was overestimated via the non-preferential model during the springs of 2016, 2017, and 2019. However, the resulting inference from the preferential model was similar to that provided in Figure \ref{fig:estimates_3} and better coincides with the removed data and scientific knowledge of the abundance process of mosquitoes; this indicates that more robust inference is obtained under the preferential model for population dynamics at the rise and decline of abundance.

\begin{figure}
    \centering
    \caption{Posterior distributions for the first day in which each species experiences population growth, $\psi$, for 2017-2019 when removing zero-count observations. Black lines depict the phenometric derived via summary statistics.}
    \includegraphics[width=\textwidth]{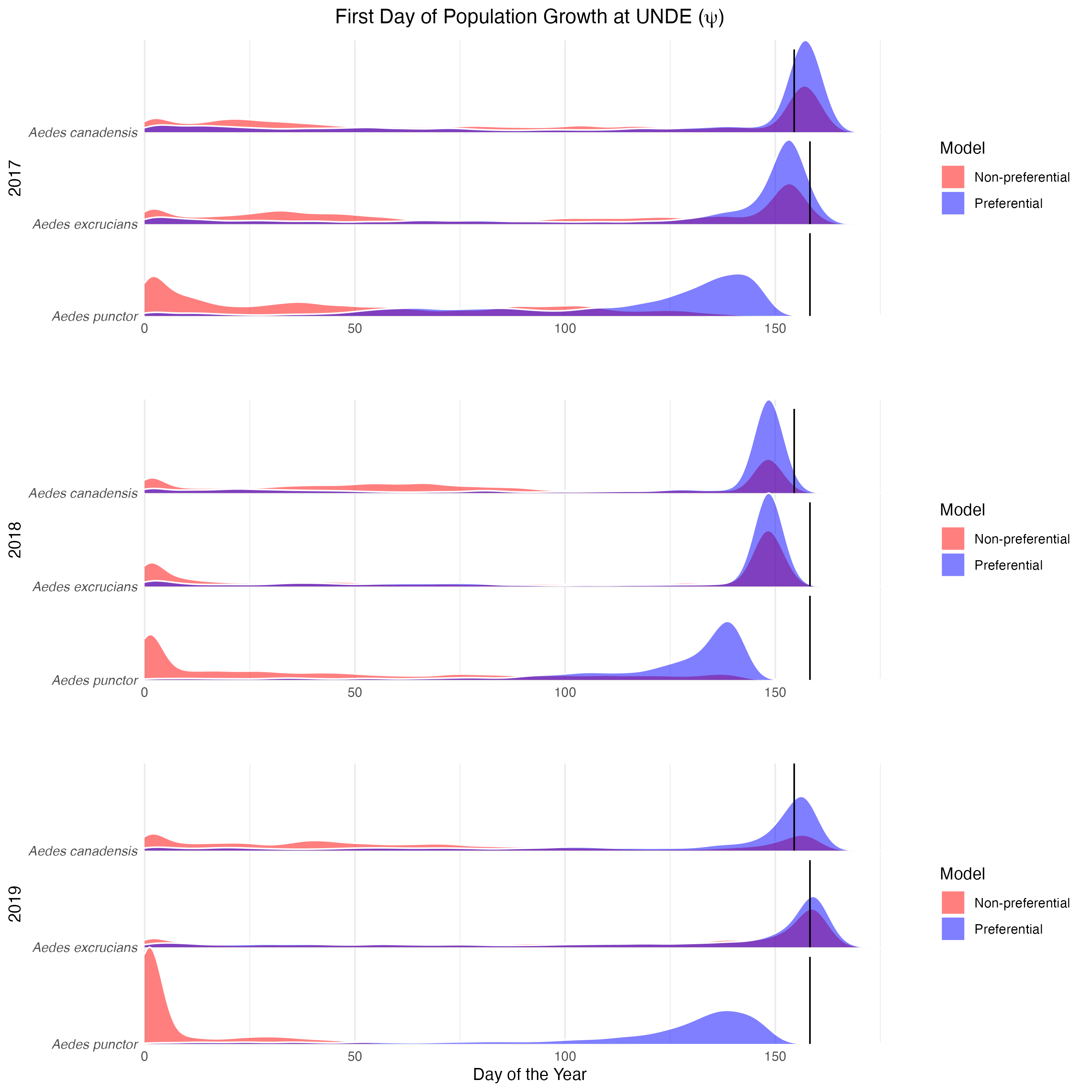}
    \label{fig:ridges-wo0}
\end{figure}

In the second scenario, the posterior inference for the phenometric $\psi$ substantially changed for the non-preferential model. As depicted in Figure \ref{fig:ridges-wo0}, the non-preferential model more heavily weighted the beginning of the year, whereas the preferential model provided similar inference as in the first scenario (see Supplemental Figure 3). The inference provided by the preferential model aligns with scientific knowledge of mosquito population dynamics under both scenarios. Posterior estimates for $\boldsymbol{\alpha}$ and $\mathbf{B}$ were inadequate under the non-preferential model in the second scenario, which significantly affected the ability to obtain robust estimates for the phenometric.

\section{Discussion}

We augmented a hierarchical abundance model to account for temporal preferential sampling and to obtain phenometrics. Temporal preferential sampling is common in abundance studies, and our extension of the model improved abundance estimates during periods of infrequent observation using the readily available data $\boldsymbol{\tau}$. Consequently, the derived phenometrics are greatly improved when accounting for preferential sampling. We derived population growth rates explicitly and inferred phenometrics by relating the latent abundance process to the stochastic Gompertz population growth function in a way that accounts for preferential sampling. The phenometrics can be compared to existing phenological studies to confirm or present new results. In addition, our model formally allowed us to quantify uncertainty associated with $\psi$, whereas most phenological approaches lack uncertainty quantification in the relevant phenometrics.

Our first case study revealed the advantages of accounting for temporal preferential sampling in the abundance model under three sampling scenarios. Under preferential switch sampling and logistic sampling, fitting the preferential model resulted in more precision when predicting unobserved abundances. In data sets with fewer, longer gaps of missing data, the proposed preferential model will outperform its non-preferential variant.

In the mosquito case study, we found that enough zero-abundance counts were recorded in the UNDE data for the non-preferential model to estimate all model parameters and, consequently, abundance. Thus, the NEON sampling mechanism did not affect posterior inference on abundance. However, the non-preferential model overestimated the growth rate during periods of infrequent observation when analyzing the subset of NEON data that only included positive counts. The second scenario is realistic because many abundance studies observe strictly positive counts. For the second scenario, the posterior inference under the non-preferential model was imprecise during the mosquito off-season, indicating that temporal preferential sampling should be accounted for when working with abundance data that are mostly positive.

During abundance study design, ecologists and wildlife biologists may apply the proposed temporal preferential sampling methods to obtain more accurate inference with less data. Researchers may allocate less resources to collect data at the start and end of the off-season and more resources during the field season when studying species with population dynamics that depend on the environment. Additionally, smaller ecological surveys may find the preferential model useful if they lack the resources to collect data consistently during an off-season.

Temporal preferential sampling has been accounted for exclusively in the continuous-time domain. Applications include monitoring air pollution \citep{shaddick2014case}, clinical trials \citep{monteiro2019modelling}, and neuroimaging \citep{poeppel2003analysis}. By contrast, we account for preferential sampling in discrete time to align with data from many abundance studies. When explicitly accounting for the dependence of observations on the process of interest, we obtain more robust inference during periods of infrequent observation.

The field of temporal preferential sampling is rapidly developing, and there is much work to be done regarding its theory and methodology. Additionally, the implementation of models that account for preferential sampling may be improved. For example, the development of algorithms that induce conjugacy for the abundance process may increase the efficiency of MCMC implementations \citep[e.g.,][]{bradley2018computationally}. Recursive and distributed computing approaches would increase the computational efficiency of temporal preferential sampling models \citep[e.g.,][]{hooten2021making}. If we seek to fit a hidden Markov model while accounting for preferential sampling, then the use of particle Gibbs filters or forward-backward algorithms may significantly improve computational efficiency \citep{beal2001infinite,tripuraneni2015particle}. Finally, interspecies interactions can be accounted in a multivariate version of our model to improve species conservation, network analysis, and community ecology.

\section*{Acknowledgements}
Data were collected by the National Ecological Observatory Network and the Oregon State PRISM Climate Group. The National Ecological Observatory Network is a program sponsored by the National Science Foundation (NSF) and operated under cooperative agreement by Battelle. This material is based in part upon work supported by the NSF through the NEON Program. Any use of trade, firm, or product names is for descriptive purposes only and does not imply endorsement by the U.S. Government.

\section*{Funding}
This research was supported by the NSF Graduate Research Fellowship Program. 

\section*{Disclosure Statement}

The authors report that there are no competing interests to declare.

\section*{Data Availability}

The datasets analyzed during this study are publicly available from the National Ecological Observatory Network (DOI: dp1.10043.001)  and PRISM climate group (https://prism.oregonstate.edu/).

\section*{Code Availability}

\textit{We will provide a GitHub link containing the code for publication.}

\newpage

\section*{Appendix A}

We solve for the species-level per capita growth rate $g_j(t)$ in (\ref{eq:growth}) and the posterior median of $g_j(t)$ in (\ref{eq:pm_growth}). From (\ref{eq:logabundance}), we can write the abundance of species $j$ on day $t$ as 
\begin{align}
    \lambda_j(t) &= \exp\left(\boldsymbol{\beta}_j(\mathbf{x}(t)-\alpha_j\mathbf{x}(t-1)) + \alpha_j\log\lambda_j(t-1) + \epsilon_{j,t}\right)\\
    &= \exp\left(\boldsymbol{\beta}_j(\mathbf{x}(t)-\alpha_j\mathbf{x}(t-1))\right)\lambda_j(t-1)^{\alpha_j}e^{\epsilon_{j,t}},
\end{align}
where $\epsilon_{j,t}\sim \text{N}(0,\sigma^2)$. If we consider the heterogeneous Malthusian growth function $\lambda_j(t)=(1+g_j(t))\lambda_j(t-1)$, we can solve for $g_j(t)$ to obtain
\begin{align}
    g_j(t) &= \frac{\lambda_j(t)}{\lambda_j(t-1)}-1\\
    &= \frac{\exp\left(\boldsymbol{\beta}_j(\mathbf{x}(t)-\alpha_j\mathbf{x}(t-1))\right)\lambda_j(t-1)^{\alpha_j}e^{\epsilon_{j,t}}}{\lambda_j(t-1)}-1\\
    &= \exp\left(\boldsymbol{\beta}_j(\mathbf{x}(t)-\alpha_j\mathbf{x}(t-1))\right)\lambda_j(t-1)^{\alpha_j-1}e^{\epsilon_{j,t}}-1.
\end{align}
Finally, $\exp(\epsilon_{j,t})$ is log-normally distributed with parameters $\mu=0$ and $\sigma^2$. Thus, $M(\epsilon_{j,t})=\exp(\mu)=1$ and
\begin{align}
    M(g_j(t)) &= \exp\left(\boldsymbol{\beta}_j(\mathbf{x}(t)-\alpha_j\mathbf{x}(t-1))\right)\lambda_j(t-1)^{\alpha_j-1}-1,
\end{align}
which is the expression in (\ref{eq:pm_growth}). We let $M(\cdot)$ denote the median of the random variable.

\newpage

\section*{Appendix B}

The full Bayesian hierarchical model that we used in our case studies is
\begin{align}
    y_j(\tilde{t}_k) &\sim \text{Pois}(\lambda_j(\tilde{t}_k)\cdot\omega(\tilde{t}_k)),\\
    \tau_i &= \begin{cases}1, & z(t_i) > 0\\ 0, & z(t_i) \le 0\end{cases},\\
    z(t_i) &\sim \text{N}\left(\theta_0 + \theta_1 \mathbbm{1}_{\left\{\boldsymbol{\lambda}(t_i)'\mathbf{1} \ge \tilde{\lambda}\right\}},1\right),\\
    \log\boldsymbol{\lambda}(t_i) &\sim \text{N}\left(\mathbf{B}\mathbf{x}(t_i) - \mathbf{A}\mathbf{B}\mathbf{x}(t_{i-1}) + \mathbf{A}\log\boldsymbol{\lambda}(t_{i-1}),\sigma^2\mathbf{I}\right), \quad i=2,...,N, \label{eq:exception}\\
    \log\boldsymbol{\lambda}(t_1) &\sim \text{N}(\boldsymbol{\mu}_1,\sigma^2_1\mathbf{I}),\\
    \boldsymbol{\beta}_j' &\sim \text{N}(\boldsymbol{\mu}_\beta,\boldsymbol{\Sigma}_\beta),\\
    \boldsymbol{\mu}_\beta &\sim \text{N}(\boldsymbol{\mu}_0,\boldsymbol{\Sigma}_0), \\
    \boldsymbol{\Sigma}_\beta &\sim \text{IW}(\boldsymbol{\Psi},\nu), \\
    \boldsymbol{\alpha} &\sim \text{N}(\boldsymbol{\mu}_\alpha,\sigma^2_\alpha\mathbf{I}), \\
    \boldsymbol{\theta} &\sim \text{N}(\boldsymbol{\mu}_\theta,\boldsymbol{\Sigma}_\theta),\\
    \sigma^2 &\sim \text{IG}(q,r), \\
    \tilde{\lambda} &\sim \text{Ga}(\alpha_\lambda,\beta_\lambda),
\end{align}
for $k=1,...,n$, $j=1,...,J$, and $i=1,...,N$ with the exception of (\ref{eq:exception}). We let $\mathbf{A}\equiv\text{diag}(\boldsymbol{\alpha})$, $\mathbf{B}\equiv(\boldsymbol{\beta}_1,...,\boldsymbol{\beta}_J)'$, and $\boldsymbol{\beta}_j\equiv(\beta_{j,1},...,\beta_{j,p})$.

\newpage

\bibliographystyle{agsm}
\bibliography{bibliography}

%\printbibliography

\newpage

%% Figure 1

%% Table 1
\begin{table}[!h]
\centering
\caption{\label{tab:RMSE} Root-mean-squared errors for the two competing models under three sampling scenarios.}
\begin{tabular}{ccc}
\multicolumn{1}{c}{\textbf{Sampling Mechanism}} & \textbf{Non-preferential Model} & \textbf{Preferential Model} \\ \hline
Random                                          & 4.393                           & \textbf{4.332}              \\
Preferential Switch                             & 12.127                         & \textbf{8.016}             \\
Logistic                                        & 8.711                          & \textbf{2.578}             
\end{tabular}
\end{table}

%% Figure 2

%% Figure 5

%% Figure 7

%% Figure 8

\end{document}